\documentclass[prl,twocolumn,aps,10pt,letterpaper, preprintnumbers,superscriptaddress]{revtex4-2}
\usepackage{upgreek}
\usepackage{verbatim}
\usepackage{amsmath}
\usepackage{latexsym}
\usepackage{revsymb}
\usepackage{yfonts}
\usepackage{ifthen}
\usepackage{tcolorbox}
\usepackage{natbib}
\usepackage{amsfonts}
\usepackage{amsmath}
\usepackage{amssymb}
\usepackage{amsthm}
\usepackage{graphicx}
\usepackage{bm}
\usepackage{bbm}
\usepackage{epsfig,color,amssymb}
\usepackage{subfigure}
\usepackage{amsfonts}
\usepackage{amscd}
\usepackage{amsmath}
\usepackage{multirow}
\usepackage{chemarrow}
\usepackage{dcolumn}
\usepackage{bm}
\usepackage{graphicx}
\usepackage{enumerate}
\usepackage{epsfig}
\usepackage{subfigure}
\usepackage{xcolor}
\usepackage{multirow}
\usepackage{ulem}
\usepackage{braket}
\usepackage{comment}
\usepackage{enumitem}
\usepackage{amsthm}
\usepackage{diagbox}
\usepackage{inputenc}
\usepackage{siunitx}
\usepackage{hyperref}

\renewcommand{\emph}[1]{\textit{#1}}

\newcommand{\ba}{\begin{array}}
\newcommand{\ea}{\end{array}}
\newcommand{\be}{\begin{equation}}
\newcommand{\ee}{\end{equation}}
\newcommand{\bea}{\begin{eqnarray}}
\newcommand{\eea}{\end{eqnarray}}

\begin{document}
\title{Geometry-Driven Lattice of Photonic Spin-Meron Tubes in Free Space}

\author{Anand Hegde}
\affiliation{Institute of Photonics Technologies, National Tsing Hua University, Hsinchu 30013, Taiwan}

\author{Komal Gupta}
\affiliation{Institute of Photonics Technologies, National Tsing Hua University, Hsinchu 30013, Taiwan}

\author{Yanan Dai}
\affiliation{Department of Physics, Southern University of Science and Technology, Shenzhen 518055, China.}

\affiliation{Quantum Science Center of Guangdong-Hong Kong-Macao Greater Bay Area , 
Shenzhen, 518045, China}

\author{Chen-Bin Huang}\email{robin@ee.nthu.edu.tw}
\affiliation{Institute of Photonics Technologies, National Tsing Hua University, Hsinchu 30013, Taiwan}

\date{\today}


\begin{abstract}

We theoretically demonstrate the first photonic spin-meron tube lattice in free space using spin-angular momentum vectors.  Square-block diffraction creates $C_{4}$-symmetric beams with $\pi/2$ phase steps. Non-paraxial spin–orbit coupling then forms forms finite-length meron tubes ($N_{sk} \approx \pm1/2$, $>25 \lambda$). Extending the formalism, we show that $C_{3}$ geometry of triangular block yields spin-skyrmion tube. Stratton-Chu theory and full-vectorial finite-difference time-domain calculations both support this material-agnostic geometric driven approach as a platform to explore the symmetry-driven free space topology.
\end{abstract}

\maketitle
Topological protection grants stability to nontrivial low-energy field configurations \cite{skyrme1961non,nagaosa2013topological}. Such protected spin textures include skyrmions and merons and appear in systems ranging from magnetism to modern optics.~\cite{fert2017magnetic,shen2024optical}. Topological photonics has grown rapidly, however, no study has realized a freely propagating lattice of spin merons in homogenous free space yet. Achieving this goal would furnish the optical counterpart of magnetic meron crystals \cite{tokura2020magnetic} along with skyrmion tubes \cite{milde2013unwinding,xing2020magnetic} and braids \cite{zheng2021magnetic}. Topological spin patterns in optics remain limited to plasmonic near fields~\cite{shi2021spin,dai2022poincare} and to polarization lattices in structured far-field beams and metasurfaces~\cite{shen2025free}.
Surface plasmon interference yields skyrmions and merons in the near field \cite{tsesses2018optical,dai2020plasmonic,ghosh2021topological} and Stokes polarization engineering produces skyrmion and meron lattices in the far field \cite{shen2022generation,bervskys2023accelerating}. All free-space demonstrations to date require bulky phase-modulation optics that create the multi-beam interference and spin–orbit coupling needed for topological control \cite{zeng2024tightly,shen2024gradient,hakobyan2025qplate}.This reliance hinders integration into compact photonic platforms. \\  

A free‐space spin‐meron lattice demands two intertwined symmetry breaking. First, continuous rotational invariance must collapse to a discrete fourfold ($\mathcal{C}_4$) symmetry which is optically realized via the coherent superposition of four non-collinear waves or vector modes (e.g., four plane waves at $\pm k\hat x,\pm k\hat y$ with successive $\pi/2$ phase shifts) creating a square array of optical vortices~\cite{lei2021photonic,marco2024propagation}. Second, a nonzero longitudinal spin component $S_z$, stemming from intrinsic spin–orbit coupling in non‐paraxial fields, is essential to tilt each vortex into a half‐skyrmion (meron)~\cite{bliokh2015spin}. Although some aspects  have been achieved, they are mostly confined to momentum‐space~\cite{guo2020momentum}, spin-angular momentum (SAM)  quasicrystals~\cite{lin2025photonic}, or focal‐region skyrmions \cite{lei2025skyrmionic}; no approach to date yields a propagation‐invariant lattice of true spin merons in homogeneous free space. This fundamental gap motivates our central question: can a geometry-imposed spin–orbit gauge field simultaneously enforce $\mathcal{C}_4$ symmetry and generate longitudinal spin component $S_z$ to realize the first propagation-invariant spin meron lattice in free-space?\\

In this work, we demonstrate a simple yet powerful method to realize stable free-space lattices of spin merons and skyrmions, overcoming the need for complex optical elements such as vortex beams or phase-engineered wavefronts. Using analytic theory and full-vectorial 3D finite-difference time-domain (FDTD) simulations, we show that illumination of a single opaque square dielectric block with circularly polarized light spontaneously generates a robust spin-meron lattice in free space. Four coherent edges enforce a discrete $\mathcal{C}_4$ symmetry and embed a fixed $\pi/2$ phase ladder, which locks a vortex lattice with a fixed $\lambda/2$ spacing. Non-paraxial spin–orbit coupling naturally induces a longitudinal spin component $S_z$, converting each optical vortex into a Bloch-type meron with half-integer skyrmion number $N_{\mathrm{sk}}\approx\pm\tfrac12$. The lattice evolves into vertical spin-meron tubes that retain their structural and topological robustness over at least $25\lambda$ of propagation, as explicitly verified by our results. Our proposed method remains valid even in realistic silicon blocks exhibiting partial absorption and internal reflections, thus highlighting its experimental viability. Generalizing the concept from meron to skyrmions, we also show that replacing the square block with a triangular polygon generates integer-charged skyrmion lattices arranged on a hexagonal grid to form free space spin-skyrmion tubes. Thus, our approach provides a geometry-driven, phase-modulation-free route to realize topological spin-texture lattices and stable spin-texture tubes directly in free space, offering broad potential for integrated photonic and topological optical applications.
\vspace{6pt}

\begin{figure*}[htbp]
    \centering
    \includegraphics[width=0.95\linewidth]{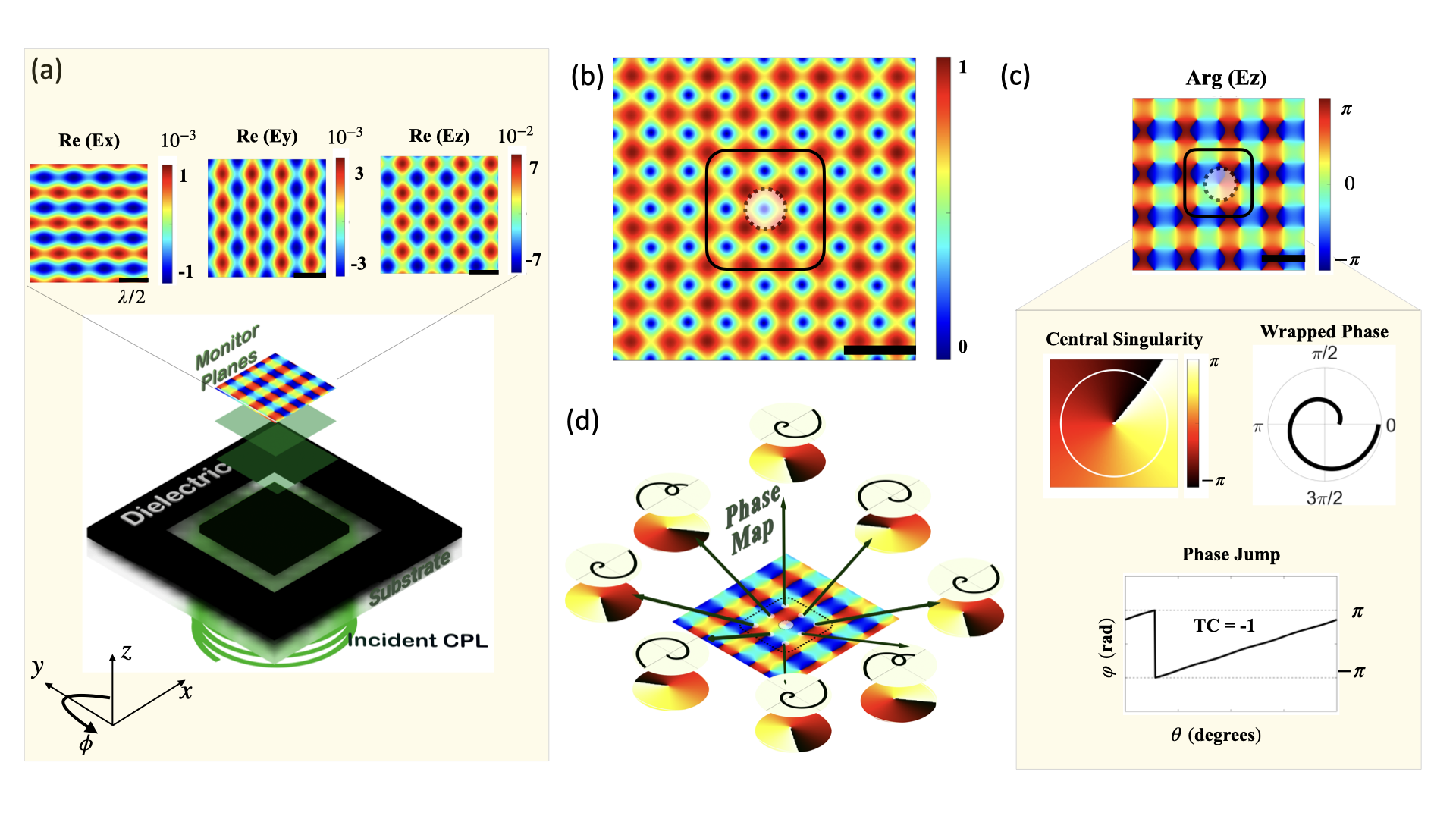}
    \caption{\textbf{Square-block diffraction from left circularly polarized light and the resulting vortex lattice.}
(a) Composite snapshot on the monitor plane ($z=2\lambda$): bottom panel sketches the opaque $64\lambda\times64\lambda$ square and the four diffracted orders (white arrows); the top row shows instantaneous real parts of $E_{x},E_{y},E_{z}$.   
(b) Normalized field magnitude $|\mathbf E|$; the nine dark minima mark vortex cores spaced by $\lambda/2$.  
(c) Wrapped phase of the longitudinal field $E_{z}$; inset plots the azimuthal phase $\varphi(\theta)$ around the central core, confirming a $2\pi$ winding and charge $m=-1$.  
(d) Phase-wheel overlay for every core: axial vortices rotate clockwise ($m=+1$), diagonal vortices counter-clockwise ($m=-1$), evidencing the fixed $\pi/2$ phase ladder enforced by the four edges.  
Scale bar: $0.5\lambda$.}
    \label{fig1corr}
\end{figure*}
A circularly polarized wave illuminating a square dielectric block breaks continuous rotational symmetry into the discrete fourfold group $\mathcal{C}_{4}$ and embeds a fixed $\pi/2$ phase ladder, thereby seeding a free-space vortex lattice. We model the block as a perfectly absorbing square of edge \(a\lambda\) with \(a\gg1\). Choosing the side length as an integral multiple of the wavelength ensures that the diffracted orders overlap coherently and form a stable pattern in a region only a few wavelengths above the center of the top face. Throughout the analysis and in Fig.~\ref{fig1corr} we set \(a=64\), while the observation (monitor) plane is placed at a height of \(z=2\lambda\). Alternative choices of \(a\) or monitor height would leave the local vortex topology and subsequent spin textures intact, provided the block size remains many wavelengths across.
The four edges serve as coherent sources whose successive $\pi/2$ phase offsets lock the lattice pitch at $\lambda/2$. The incident circularly polarized wave is
\begin{equation}
\mathbf E_{\text{inc}} = E_{0}\bigl(\hat{\mathbf x}\pm i\hat{\mathbf y}\bigr)e^{ikz},
\qquad
\mathbf H_{\text{inc}} = \frac{E_{0}}{\eta_{0}}\bigl(\hat{\mathbf y}\mp i\hat{\mathbf x}\bigr)e^{ikz},
\label{eq:inc}
\end{equation}
with \(k = 2\pi/\lambda\) and free-space impedance \(\eta_{0}\).\\

We model the diffracted field in the half-space \(z>0\) using the Stratton–Chu formalism~\cite{stratton1939diffraction}, which expresses every component of \(\mathbf E\) in terms of the tangential incident fields on the aperture plane. Because real absorbers possess finite thickness, their sidewalls support additional diffraction currents. These are accounted for by explicitly retaining the magnetic line-integral (edge) term that runs around the perimeter, as detailed in Supplementary Section~S1. This contour-augmented formulation preserves the fourfold phase ladder and longitudinal spin injection observed in numerical simulations, while avoiding full volumetric complexity. For each Cartesian component \(i\in\{x,y,z\}\), the diffracted field is
\begin{widetext}
\begin{equation}
E_{i}^{\text{diff}}(\mathbf r')
  = \frac{1}{4\pi}\iint_{S}
      \Bigl[
        i\omega\mu_{0}\,(\mathbf n\times\mathbf H)_{i}\,G
        + (\mathbf n\times\mathbf E)_{j}\,\partial_{k}G
        - (\mathbf n\times\mathbf E)_{k}\,\partial_{j}G
      \Bigr]\mathrm dS
    \;-\;
    \frac{1}{4\pi i\omega\varepsilon_{0}}
    \oint_{C}(\mathbf H\!\cdot\!\mathrm d\boldsymbol\ell)\,\partial_{i}G,
\label{eq:SC}
\end{equation}
\end{widetext}
where the cyclic permutation \((i,j,k)=(x,y,z)\!\to\!(y,z,x)\!\to\!(z,x,y)\) enforces right-handed order. The scalar Green function is \(G = e^{ik|\mathbf r'-\mathbf r|}/|\mathbf r'-\mathbf r|\) and the outward unit normal on the aperture is \(\mathbf n = -\hat{\mathbf z}\). For the square block, the contour \(C\) lies entirely in the \(x\!-\!y\) plane, and thus the edge term contributes primarily to the longitudinal field \(E_{z}\). All integrands vanish within the perfectly absorbing interior, requiring no fitting parameters. A detailed derivation of this analytical formulation, along with benchmarking against FDTD simulations, is presented explicitly in Supplementary Section~S1.\\

Fig.~\ref{fig1corr}(a) compiles analytical Stratton–Chu results for a left circularly polarized (LCP) illumination at \(z=2\lambda\). The bottom vignette illustrates the square absorber and its four diffracted fields, while the three top panels show instantaneous real parts \(\operatorname{Re}\{E_{x}\}\), \(\operatorname{Re}\{E_{y}\}\), and \(\operatorname{Re}\{E_{z}\}\) at arbitrary instances. The complete corresponding  magnetic fields derived via Maxwell’s curl relation are provided in Supplementary Section~S2.\\

Panel~\ref{fig1corr}(b) shows the normalized electric-field magnitude \(|\mathbf E|\), where dark minima correspond to candidate vortex cores. Panel~\ref{fig1corr}(c) shows the wrapped phase of the longitudinal component \(E_{z}\), revealing clear phase discontinuities, especially prominent at the block center \((x,y)=(0,0)\). The inset explicitly plots the azimuthal phase \(\varphi(\theta)\) around this core, confirming a full \(2\pi\) winding indicative of a topological charge \(m=-1\). The accompanying video of the \(E_{z}\) field vividly demonstrates the local revolution of the field around these singularities. The four equivalent edges enforce fixed \(\pi/2\) phase offsets, thus the transverse field near a vortex core adopts the Archimedean spiral form:
\begin{subequations}\label{eq:spiral}
\begin{align}
E_{\perp}(r,\theta) &= A\,r\,e^{\,i(\theta+\pi/2)}, \\
\varphi(r,\theta)   &= \theta + k r.
\end{align}
\end{subequations}
Within the \(\lambda\times\lambda\) box highlighted in Fig.~\ref{fig1corr}(b-d), eight additional vortex cores flank the central singularity. Their locations match precisely the analytical zeros of the longitudinal field,
\begin{equation}
E_{z}(x,y)\sim B\,\sin(kx)\cos(ky).
\label{eq:Ezzeros}
\end{equation}

\paragraph{Axial nodes:} \(\sin(kx)=0\) and \(\cos(ky)\neq0\) yield four Cartesian points at \(x=\pm\Delta\), \(y=0\), with \(\Delta=\lambda/2\).
\paragraph{Diagonal nodes:} \(\cos(ky)=0\) and \(\sin(kx)\neq0\) yield four points at \(x=\pm\Delta\), \(y=\pm\Delta\).

Integration of the transverse phase \(\varphi=\arg(E_x+iE_y)\) around small loops enclosing each node gives the topological charges,
\begin{equation}
m=\frac{1}{2\pi}\oint_{\mathcal C}\nabla\varphi\!\cdot d\boldsymbol\ell.
\label{eq:charge}
\end{equation}

The fixed \(\pi/2\) phase ladder ensures axial vortices have charge \(m=+1\), while diagonal vortices carry \(m=-1\), thus forming a robust square vortex lattice. This vortex distribution evolves slightly at different propagation distances due to varying relative phases of scattered fields from different edges; nonetheless, as detailed in Supplementary Section~S3 and accompanying videos, the global spin textures formed initially remain robust and propagate over extended distances without significant distortion, a key point explored further below.\\
\begin{figure}[t!]
\includegraphics[width=\linewidth]{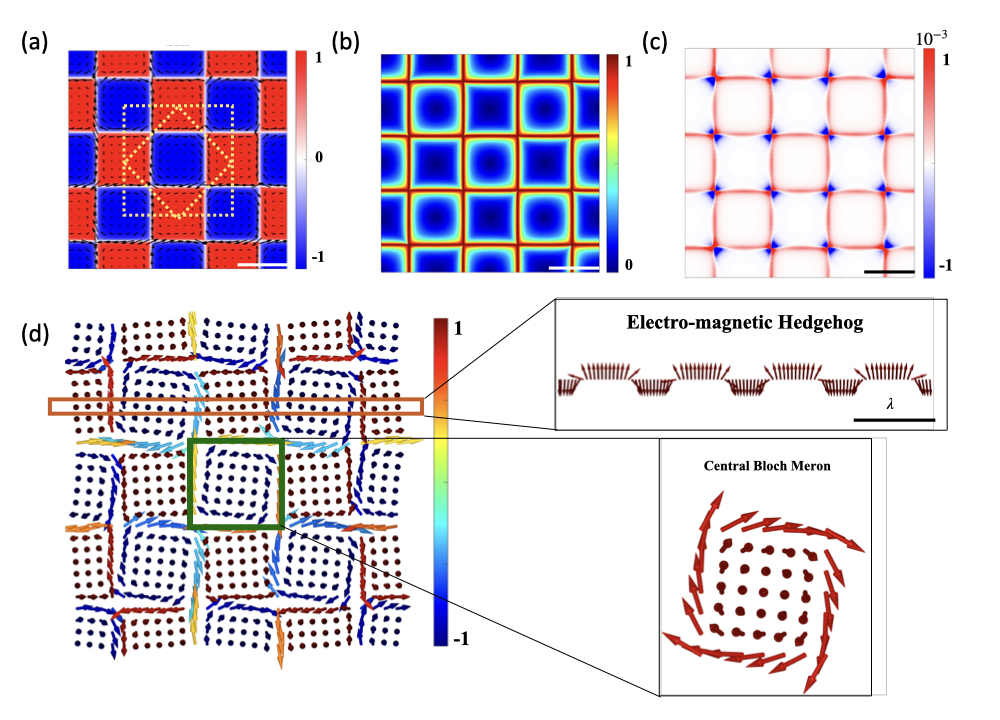}
\caption{\label{fig2} {\bf Formation of free-space merons at an arbitrary monitor height of  $2\lambda$.} (a) Normalized out-of-plane component of the local electromagnetic SAM, arrows indicate the direction of the normalized in-plane SAM. The dotted lines show the smallest square and diamond unit cells. (b) Normalized magnitude of the in-plane spin component. (c) Skyrmion density. (d) 3D spin texture with a robust hedgehog structure and a central meron of the Bloch type.}
\end{figure}

Since the four diffracted fields arrive with a fixed \(\pi/2\) phase ladder, moving one quadrant in the \((x,y)\)-plane always advances the wrapped phase by \(\pi/2\). This effectively results in a clockwise Archimedean spiral (\(m=+1\)) for each axial node, while the diagonal nodes inherit a counterclockwise spiral (\(m=-1\)). Fig.~\ref{fig1corr}(d) clearly visualizes this alternation, with axial phase wheels rotating in one direction and diagonal ones in the opposite, separated by exactly \(\pi/2\). Summing the nine singularities within one lattice cell—central (\(m=-1\)), four axial (\(+1\)), and four diagonal (\(-1\))—gives
\[
\sum m = -1 + 4(+1) + 4(-1) = -1,
\]
matching the helicity of the incident left-circularly polarized beam. Orbital angular momentum is thus globally conserved even though local vortex signs alternate. The resulting \(3\times3\) matrix of vortices, spaced by \(\Delta=\lambda/2\) in both \(x\) and \(y\), forms a propagation-invariant square vortex lattice that serves as the basis for the richer spin textures analyzed next.\\

We evaluate the spin texture numerically using a nanometer-meshed 3D FDTD simulation. Detailed simulation parameters, boundary conditions are provided in Supplementary Section~S2. From the simulated electric and magnetic fields we compute the total SAM density~\cite{bliokh2014extraordinary},
\begin{equation}
\mathbf S(\mathbf r)=
\frac{\varepsilon_{0}}{4\omega}\,
\operatorname{Im}\!\bigl(\mathbf E^{*}\!\times\!\mathbf E\bigr)
+
\frac{\mu_{0}}{4\omega}\,
\operatorname{Im}\!\bigl(\mathbf H^{*}\!\times\!\mathbf H\bigr),
\label{eq:SAM}
\end{equation}
where \(\omega\) is the optical frequency.


In Fig.~\ref{fig2}(a), the normalized longitudinal spin component 
\(S_{z}\) is shown in color—red for \(+1\) and blue for \(-1\). Arrows
represent the normalized in-plane spin vector \(\mathbf
S_{\!\parallel}/|\mathbf S|\). Dashed lines outline the smallest square
and diamond unit cells. Fig.~\ref{fig2}(b) presents the normalized
magnitude \(|\mathbf S_{\!\parallel}|/|\mathbf S|\), which peaks along
lattice links and falls to zero at vortex cores, emphasizing the
dominance of in-plane spin away from singularities. A Bloch-type meron
occurs when the circulating azimuthal spin density and longitudinal
component \(S_{z}\) satisfy the condition that the ratio \(|\mathbf 
S_{\!\parallel}| /S_{z}\) rises from zero at the core to a maximum on
an annular ring and decays thereafter. This ensures \(\mathbf
S_{\!\parallel}\) remains tangential, completing a full \(2\pi\)
rotation without radial contributions, while \(S_{z}\) smoothly
vanishes at the core without changing sign. \\

 The corresponding skyrmion-density map, defined as
\begin{equation}
\rho_{\mathrm{sk}}=
\frac{1}{4\pi}\,
\mathbf S\cdot\bigl(\partial_{x}\mathbf S\times\partial_{y}\mathbf S\bigr),
\label{eq:rho}
\end{equation}
appears in Fig.~\ref{fig2}(c). High density at vortex positions indicates meron formation. 
\begin{table}[h]
  \caption{Skyrmion number per cell for different boundary shapes and lattice sizes.}
  \label{tab1}
  \begin{ruledtabular}
    \begin{tabular}{lcc}
      Shape of the boundary & Number of cells & $N_{\mathrm{sk}}$ / site \\
      \hline
      Square  &  4  & 0.46 \\
        &  9  & 0.51 \\
       & 16  & 0.48 \\
       \hline
      Diamond &  2  & 0.48 \\
       &  8  & 0.47 \\
       & 18  & 0.48 \\
    \end{tabular}
  \end{ruledtabular}
\end{table}
Integrating \(\rho_{\mathrm{sk}}\) over square and diamond unit cells of varying size (Table \ref{tab1}) consistently gives skyrmion numbers near \(N_{\mathrm{sk}}\approx\frac12\), confirming the half-integer nature of each lattice cell. Fig.~\ref{fig2}(d) demonstrates that the field profiles satisfy these criteria at each lattice site. The cross-sectional “electromagnetic hedgehog” structure clearly shows the gradual tilt. The inset highlights the central Bloch-type meron. Supplementary Section~S3 and accompanying videos demonstrate explicitly the stability and persistence of these spin textures over the full optical pulse duration. The results presented thus firmly establish the emergence of robust, geometry-driven Bloch-type meron lattices in free space.\\

\begin{figure}[h!]
    \centering
    \includegraphics[width=\linewidth]{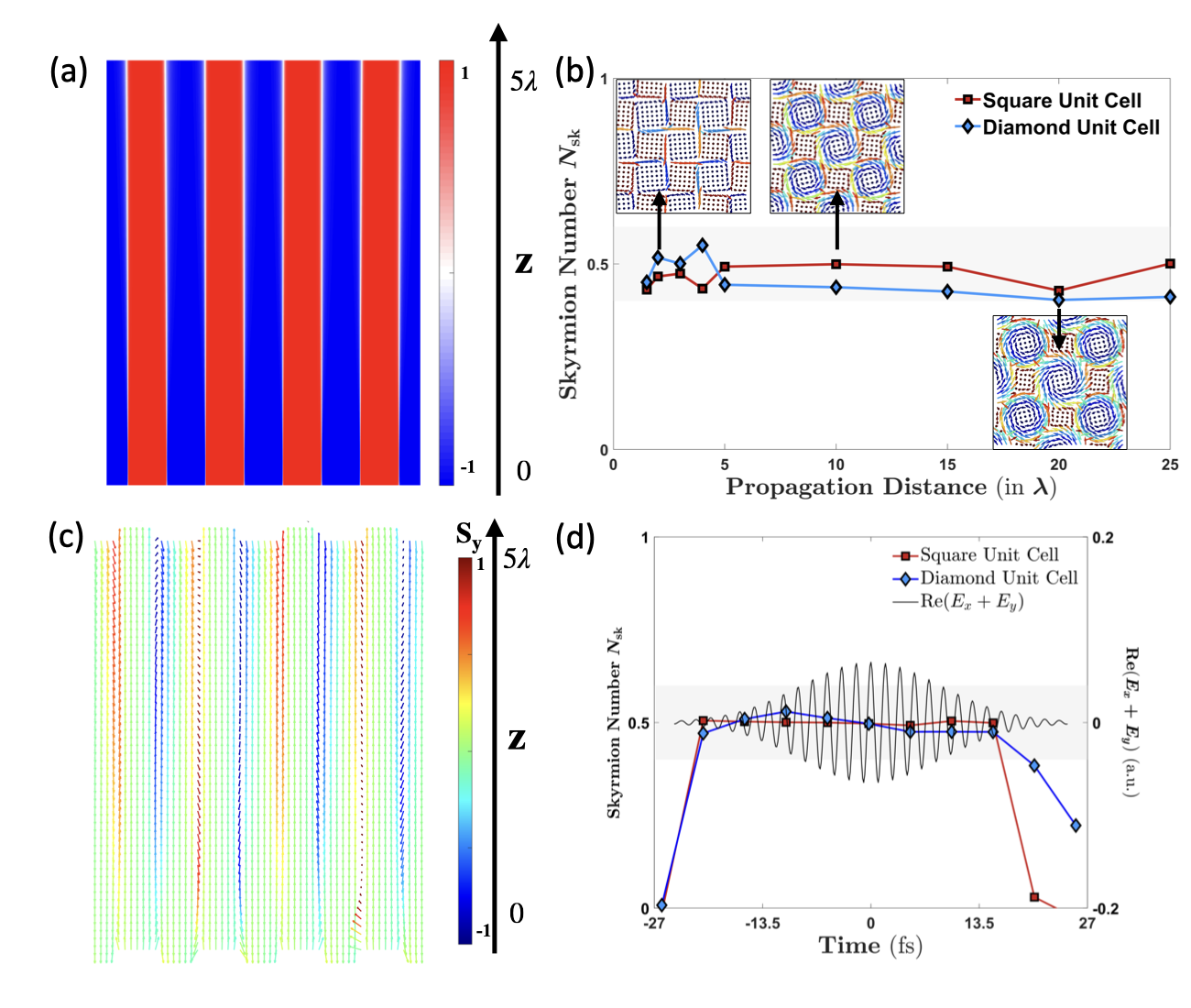}
    \caption{{\bf Spin–meron tubes in free space:} (a) Normalized longitudinal spin $S_z(x,z)$ on an $x$–$z$ slice; straight color bands indicate negligible lateral drift and delineate a tube lattice. (b) Skyrmion number $N_{\mathrm{sk}}$ versus propagation distance above the block for square and diamond unit cells. Inset: three-dimensional spin texture at observation planes $z/\lambda=1.5,\,10,\,20$. (c) Lateral 3D view of the spin field revealing vertical Bloch–type meron tubes. (d) Skyrmion number for square and diamond unit cells versus time and in-plane field $\mathbf{E}_{\parallel}$ during the optical pulse.}
    \label{fig3}
\end{figure}

Having established the Bloch‑type spin-meron lattice on a single plane
in free space, we now track its evolution in propagation and time,
summarized in Fig. 3. Panel (a) plots the normalized longitudinal spin
$S_z(x,z)$ on an $x–z$ slice. The straight, vertically extended color
bands show that the vortex cores form a tube lattice with negligible
lateral drift as $z$ increases. Panel (b) quantifies the topology: the
skyrmion number $N_\mathrm{sk}$, evaluated over both square and diamond unit cells, is preserved with distance once the field is in the radiative zone. Insets display the full three‑dimensional spin texture at $z/\lambda = 1.5, 10, \; \mathrm{and} \; 20$, confirming that the meron charge and helicity patterns are preserved during free‑space propagation, aside from a gradual amplitude reduction. Panel (c) provides an lateral 3D rendering of spin vectors in $x-z$ cross-sectional plane, revealing vertical spin-meron tubes in free space analogs of magnetic skyrmion strings extending over many wavelengths without detectable bending or breakup \cite{milde2013unwinding,zheng2021magnetic}.
Finally, panel (d) shows the stability of quasiparticles by following $N_\mathrm{sk}$ through the optical pulse at the height of $z=20\lambda$. The skyrmion number, computed concurrently with the in‑plane electric field $\mathbf{E}_{\parallel}$, rises from zero as the pulse builds, plateaus near its steady value during the peak, and returns toward zero as the field vanishes, demonstrating that the meron lattice is formed and sustained by the transient vectorial field.\\

To further validate the robustness, Supplementary Section~S3 explicitly verifies that these tubes persist at least up to \(25\lambda\). At larger propagation distances, central and diagonal merons expand slightly, exhibiting increased twisting, while axial merons remain highly localized and robust. Despite these minor deformations, all meron tubes preserve their overall lattice arrangement, helicities and topological integrity throughout extended propagation. Ultimately, the propagation distance of this lattice is limited not by lateral drift but by the breakdown of the non-paraxial regime itself, conveniently characterized by the Fresnel number,
\begin{equation}
N_{F}=\frac{a^{2}}{\lambda\,z},
\label{eq:FresnelDef}
\end{equation}
where \(a\) is the side length of the aperture. For our \(64\lambda\times64\lambda\) square (\(a=64\lambda\)), the most distant plane numerically studied (\(z=25\lambda\)) still yields a large Fresnel number,
\begin{equation}
N_{F}=\frac{(64\lambda)^{2}}{\lambda\,(25\lambda)}
      \approx 164,
\label{eq:FresnelValue}
\end{equation}
well inside the non-paraxial domain. Thus, the lattice could in principle remain intact up to several hundred wavelengths before entering the paraxial regime.

\begin{figure}[t!]
\includegraphics[width=\linewidth]{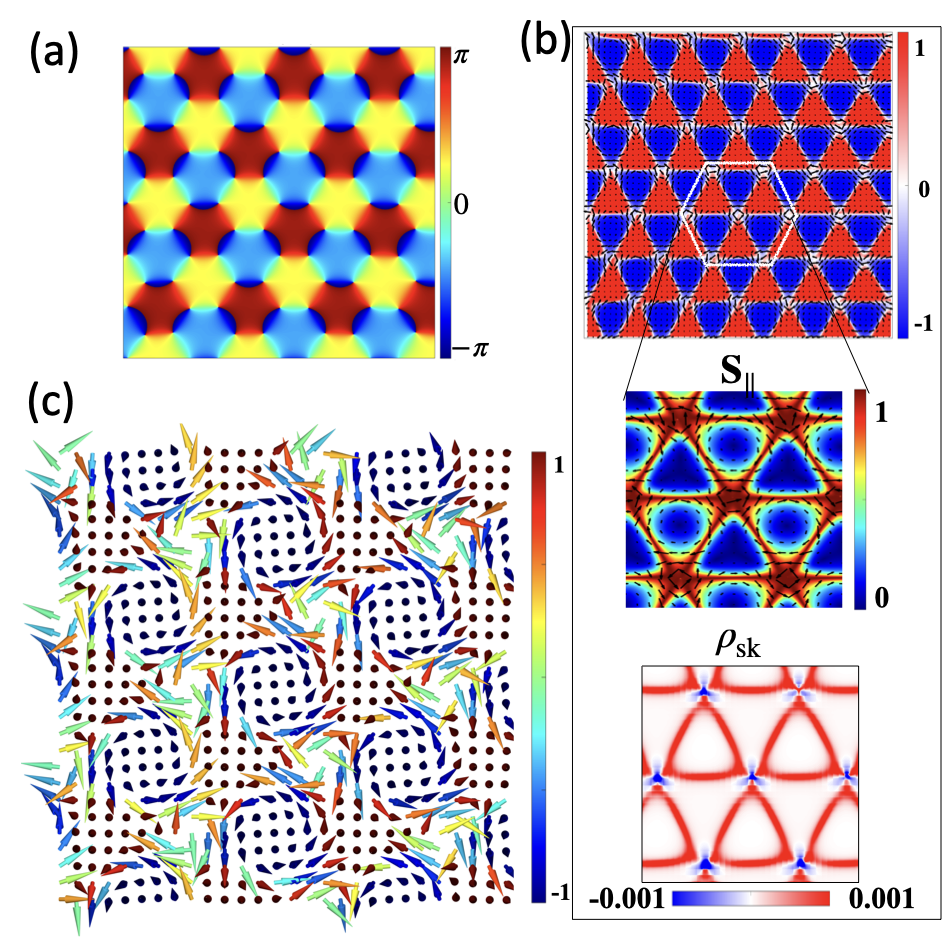}
\caption{\textbf{Hexagonal skyrmion lattice from a triangular block.}
(a) Wrapped phase of the longitudinal field \(E_z\); the \(2\pi/3\)
phase ladder set by the three edges produces a honeycomb of vortices.
(b) Longitudinal spin \(S_z\) (color) and in-plane spin
\(\mathbf S_\parallel/|\mathbf S|\) (arrows); dashed lines outline a
hexagonal unit cell. Insets: in-plane spin magnitude
\(|\mathbf S_\perp|\) and skyrmion-density map within one cell.
(c) Cross-sectional view of the 3-D spin field showing vertical skyrmion tubes on a hexagonal lattice. Scale bar: \(0.5\lambda\).}
\label{fig4}
\end{figure}

Replacing the square absorber with an \emph{equilateral triangle} of edge \(64\lambda\) (centered at \(z=0\)) swaps the fourfold ladder for a threefold one. The three equal edges launch diffracted orders separated by \(120^{\circ}\), with relative phases \(0,\,2\pi/3,\,4\pi/3\). This fixed \(C_{3}\) phase ladder seeds the hexagonal vortex array observed in the wrapped longitudinal phase of Fig.~\ref{fig4}(a). Here \(m=\pm1\) singularities form a honeycomb lattice, with phase discontinuities locked to the edge bisectors.

Evaluating the full SAM density using Eq.~\eqref{eq:SAM} yields the spin textures shown in Fig.~\ref{fig4}(b). The color shows \(S_{z}\), arrows represent the normalized in-plane spin \(\mathbf S_{\!\parallel}/|\mathbf S|\), and the dashed hexagon outlines the primitive cell. Integrating the skyrmion-density map \(\rho_{\mathrm{sk}}\) (lower inset) over one hexagonal unit cell yields an integer skyrmion number \(N_{\mathrm{sk}}=\pm1\), contrasting the half-integer merons of the square lattice. This integer charge results directly from the \(C_{3}\) symmetry. Fig.~\ref{fig4}(c) explicitly shows the hexagonal lattice of 3D spin textures which form lattice of spin-skyrmion tubes in free space, analogous to the square-lattice meron tubes in Supplementary Section~S3.\\

The geometric origin of this phenomenon is material-agnostic, as explicitly confirmed numerically using a realistic silicon block with partial absorption and multiple internal reflections (see Supplementary Section~S4). In this practical scenario, meron tubes remain robust but display slight deformations due to additional scattering and higher-order mode overlaps. These two examples illustrate a general principle: an opaque polygon with \(C_{n}\) symmetry enforces an \(n\)-order phase ladder and thus forms robust, propagating spin meron or skyrmion lattices and their by respective tubes in free space.\\

In summary, we have shown that edge diffraction from a single opaque polygon converts a circularly polarized beam into a self-organized lattice of spin textures in free space. A \(64\lambda\) square delivers a \(C_{4}\) \(\pi/2\) phase ladder that locks a \(\lambda/2\)-spaced vortex array and evolves into Bloch-type meron tubes with half-integer charge, stable for at least \(25\lambda\) of propagation. Swapping to a triangular block replaces the ladder with \(C_{3}\) symmetry and yields an integer skyrmion lattice on a hexagonal grid. Any \(C_{n}\) polygon therefore generates an \(n\)-gon topological lattice through geometry alone, a mechanism that remains robust even in realistic silicon block with partial absorption. Since our approach only requires opaque edges, such scheme could be etched directly onto photonic chips or scaled across the electromagnetic spectrum, offering a compact route to three-dimensional topological light, spin-orbit photonic logic, and resilient optical information channels.

%
%
%
%
%
%

%
%
%
%

%
%
%
%

%
%
%

%
%




\begin{acknowledgments}
This work was supported by the National Science and
Technology Council, Taiwan under Grants 113-2221-E-007-066-MY2  and 114-2112-M-007-007-. 
\end{acknowledgments}

\section*{Data Availability Statement}
The data that support the findings of this study are available from the corresponding author upon reasonable request.

\bibliographystyle{apsrev4-1}

\bibliography{apssamp}

\end{document}


\title{Supplementary Information}

\author{Anand Hegde}
\affiliation{Institute of Photonics Technologies, National Tsing Hua University, Hsinchu 30013, Taiwan}

\author{Komal Gupta}
\affiliation{Institute of Photonics Technologies, National Tsing Hua University, Hsinchu 30013, Taiwan}

\author{Yanan Dai}
\affiliation{Department of Physics, Southern University of Science and Technology, Shenzhen 518055, China.}

\affiliation{Quantum Science Center of Guangdong-Hong Kong-Macao Greater Bay Area , 
Shenzhen, 518045, China}

\author{Chen-Bin Huang}\email{robin@ee.nthu.edu.tw}
\affiliation{Institute of Photonics Technologies, National Tsing Hua University, Hsinchu 30013, Taiwan}

\date{\today}




\maketitle

\section*{S1. Stratton--Chu Model of Diffraction of a Rectangular Block}

We analyze the diffraction of a normally incident, right-circularly-polarized (RCP) plane wave by a finite square dielectric absorber occupying \(-a\le x,y\le a\) in the plane \(z=0\) using Stratton-Chu formulation \cite{stratton1939diffraction}.  
The fictitious absorber has complex index \(\tilde n = n+i\kappa = 4+5i\), giving the Fresnel coefficient for normal incidence  
\[
r=\frac{1-\tilde n}{1+\tilde n}= -0.8-0.2\,i .
\]

The incident fields in vacuum (\(z<0\)) are  
\[
\mathbf E^{\text{inc}} = E_0(\hat{\mathbf x}+i\hat{\mathbf y})e^{ikz},\qquad
\mathbf H^{\text{inc}} = \frac{E_0}{\eta_0}(-i\hat{\mathbf x}+\hat{\mathbf y})e^{ikz},
\]
with \(k=2\pi/\lambda\) and \(\eta_0\simeq 377\;\Omega\). 

\vspace{4pt}
\noindent\emph{Stratton–Chu representation.}  
For the vacuum half-space \(z>0\) the diffracted field is  
\[
\mathbf{E}^{\text{diff}}(\mathbf r')=
\frac{1}{4\pi}\iint_{S}
      \!\Bigl[i\omega\mu_0(\mathbf n\times\mathbf H)G
             +(\mathbf n\times\mathbf E)\times\nabla G\Bigr]dS
-\frac{1}{4\pi i\omega\varepsilon_0}
      \oint_{C}(\mathbf H\!\cdot d\boldsymbol\ell)\,\nabla G,
\]
where \(G=e^{ikR}/R\), \(R=|\mathbf r'-\mathbf r|\), and the downward normal is \(\mathbf n=-\hat{\mathbf z}\).

\vspace{4pt}
\noindent\emph{Boundary fields.}  
Immediately above the absorber (\(z=0^{+}\))
\[
\mathbf E_t=(1+r)E_0(\hat{\mathbf x}+i\hat{\mathbf y}),\qquad
\mathbf H_t=(1-r)\frac{E_0}{\eta_0}(-i\hat{\mathbf x}+\hat{\mathbf y}),
\]
so the surface integrand becomes
\[
\mathbf f^{(S)}=
i\omega\mu_0\frac{(1-r)E_0}{\eta_0}G
           \begin{pmatrix}1\\ i\\ 0\end{pmatrix}
\;+\;
(1+r)E_0
           \begin{pmatrix}
              -\partial_z G\\ -i\partial_z G\\ \partial_x G+i\partial_y G
           \end{pmatrix}.
\]

Along the square boundary \(C=\partial[-a,a]^2\) the contour integrand is
\[
\mathbf f^{(C)}=
-\frac{\eta_0(1-r)E_0\,h_m}{ik}
          \Bigl(ik-\frac{1}{R}\Bigr)\frac{e^{ikR}}{R^{3}}
          \begin{pmatrix}x'-x\\ y'-y\\ z'-z\end{pmatrix}d\ell ,
\]
with \(h_m=i,1,-i,-1\) for the four edges traversed clockwise.

\vspace{4pt}
\noindent\emph{Explicit Cartesian components.}  
Substituting the boundary fields into the Stratton–Chu formula yields

\[
\boxed{%
\begin{aligned}
E_x^{\text{diff}}(\mathbf r') &=
\frac{1}{4\pi}\iint_S\Bigl[
      i\omega\mu_0(\mathbf n\times\mathbf H)_x\,G
     -(\mathbf n\times\mathbf E)_z\,\partial_y G
     +(\mathbf n\times\mathbf E)_y\,\partial_z G
\Bigr]dS
-\frac{1}{4\pi i\omega\varepsilon_0}\oint_C(\mathbf H\!\cdot d\boldsymbol\ell)\,\partial_xG,\\[6pt]
E_y^{\text{diff}}(\mathbf r') &=
\frac{1}{4\pi}\iint_S\Bigl[
      i\omega\mu_0(\mathbf n\times\mathbf H)_y\,G
     -(\mathbf n\times\mathbf E)_x\,\partial_z G
     +(\mathbf n\times\mathbf E)_z\,\partial_x G
\Bigr]dS
-\frac{1}{4\pi i\omega\varepsilon_0}\oint_C(\mathbf H\!\cdot d\boldsymbol\ell)\,\partial_yG,\\[6pt]
E_z^{\text{diff}}(\mathbf r') &=
\frac{1}{4\pi}\iint_S\Bigl[
      i\omega\mu_0(\mathbf n\times\mathbf H)_z\,G
     +(\mathbf n\times\mathbf E)_x\,\partial_y G
     -(\mathbf n\times\mathbf E)_y\,\partial_x G
\Bigr]dS
-\frac{1}{4\pi i\omega\varepsilon_0}\oint_C(\mathbf H\!\cdot d\boldsymbol\ell)\,\partial_zG .
\end{aligned}}
\]

For on-axis observation (\(x'=y'=0\)) symmetry forces \(E_x^{\text{diff}}=E_y^{\text{diff}}=0\) and the remaining integrals cancel exactly, giving \(E_z^{\text{diff}}(0,0,z')=0\).  These properties have been verified analytically and numerically, and the divergence-free condition \(\nabla' \!\cdot\!\mathbf E^{\text{diff}}=0\) is satisfied identically.\\

\begin{figure}[htbp]
    \centering
    \includegraphics[width=0.8\linewidth]{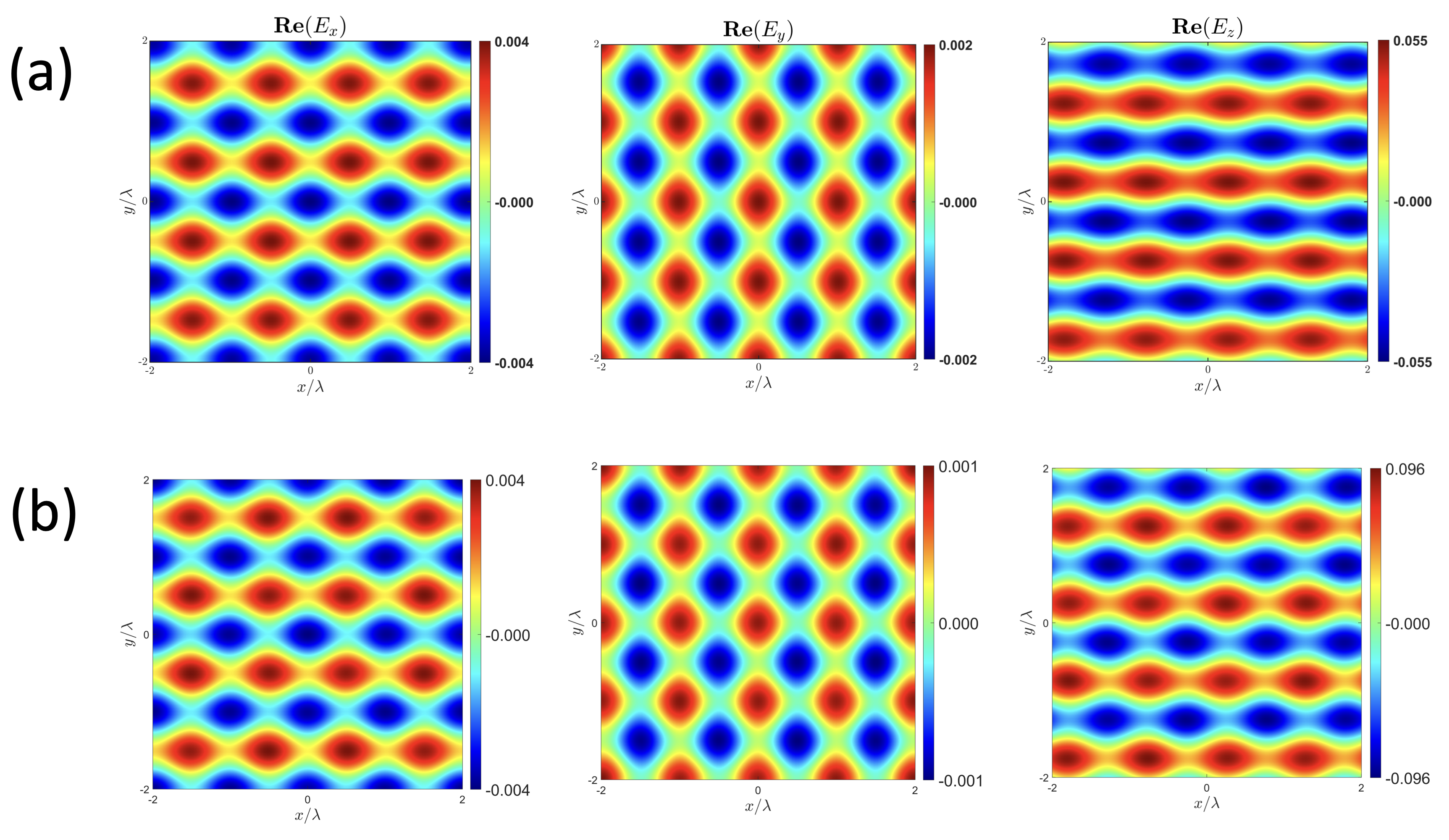}
    \caption{Comparison between FDTD simulations and the analytical Stratton–Chu model for an opaque square absorber (\(40\lambda\times40\lambda\)) illuminated by LCP light, monitored at \(z=2\lambda\). (a) FDTD simulation result, (b) analytical model. The field distributions show excellent agreement, with minor amplitude differences owing to absorber thickness and resolution variations.}
    \label{fig:S1}
\end{figure}

We validate our analytical Stratton–Chu model against 3-D finite-difference time-domain (FDTD) simulations. Fig.~\ref{fig:S1} compares the purely diffracted electric-field distributions obtained from both methods for an opaque absorber of size \(40\lambda \times 40\lambda\), illuminated by left circularly polarized (LCP) light, evaluated at a height of \(2\lambda\). Panel (a) shows the FDTD result, while panel (b) is the analytical calculation. The two field patterns exhibit excellent qualitative agreement, with identical vortex lattice positions and symmetry. A minor amplitude difference between the two results is attributed to the finite thickness of the dielectric absorber included in the FDTD model, as well as resolution differences inherent in numerical discretizations versus analytical integration. This comparison confirms the reliability of the analytical method as a robust approximation of the full 3-D problem.

\section*{S2. Magnetic field components and time evolution of spin textures}

\begin{figure}[htbp]
    \centering
    \includegraphics[width=\linewidth]{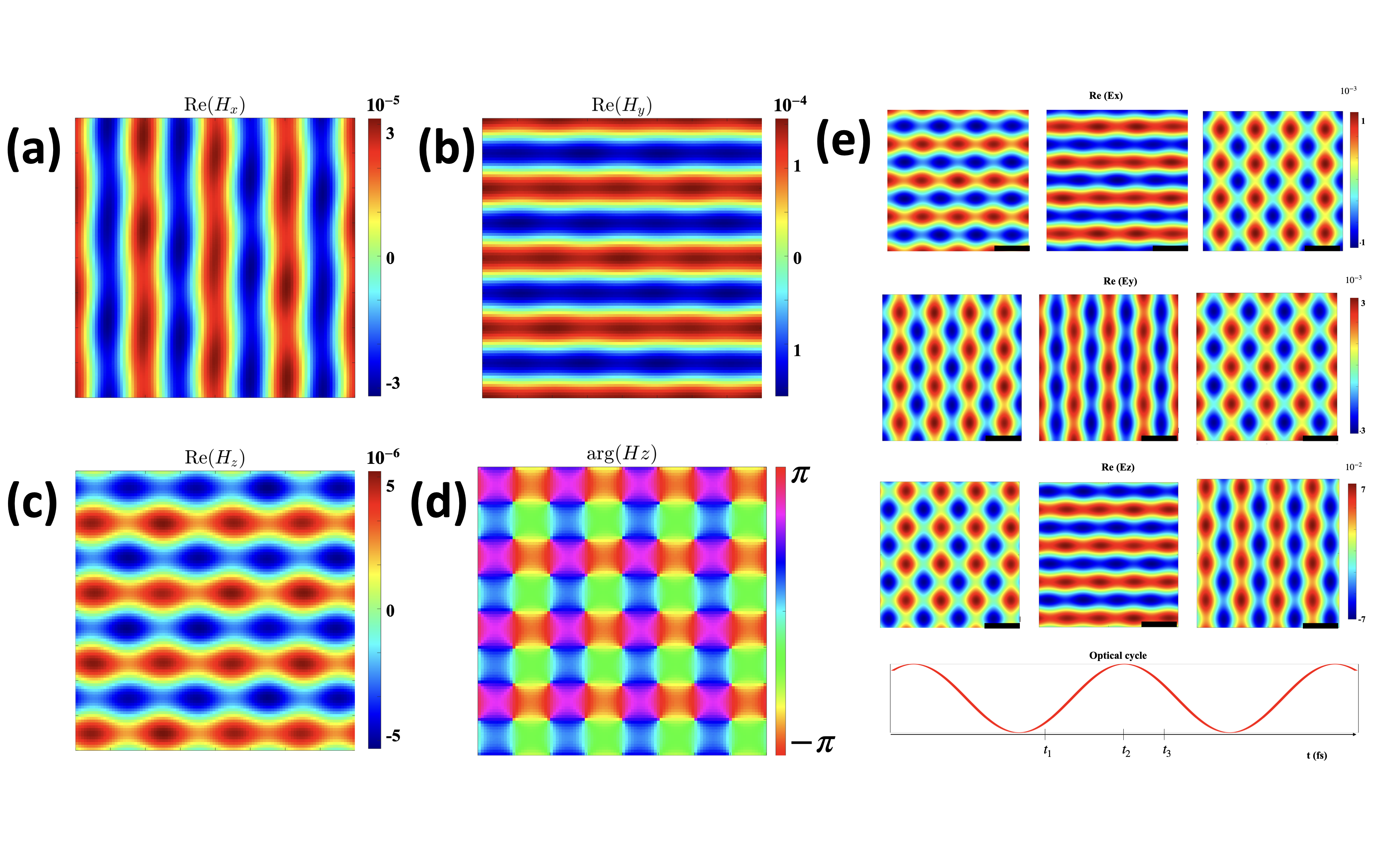}
    \caption{Magnetic field components and temporal evolution of fields. (a--c) Real parts of \(H_x\), \(H_y\), and longitudinal \(H_z\) components at the monitor plane (\(z=2\lambda\)). (d) Wrapped phase distribution of the longitudinal component \(H_z\). (e) Time evolution of electric field components \(E_x\), \(E_y\), and \(E_z\) across three arbitrary instances \(t_1\), \(t_2\), and \(t_3\) within one optical cycle. The accompanying videos clearly visualize the local rotation of \(E_z\) about vortex cores and demonstrate the robust stability of the spin texture and skyrmion density throughout the pulse length. Scale bars: \(0.5\lambda\).}
    \label{fig:S2}
\end{figure}

We further calculated the magnetic field components from the electric fields using Maxwell's curl equations (\(\mathbf H = (i/\omega\mu_0)\nabla\times\mathbf E\)). Fig.~\ref{fig:S2} shows the magnetic field distributions obtained for the same case as above. Panels (a)--(c) show the instantaneous real parts of \(H_x\), \(H_y\), and \(H_z\), respectively. While the magnitudes of the magnetic fields are significantly smaller than those of the electric fields, they still exhibit rich phase structures and a notably non-vanishing longitudinal component \(H_z\). Panel (d) shows the wrapped phase of the longitudinal magnetic field component \(H_z\), revealing the complex and ordered arrangement of singularities.

Panel (e) shows snapshots of the electric field components \(E_x\), \(E_y\), and \(E_z\) at three arbitrary time instances (\(t_1\), \(t_2\), \(t_3\)) within one optical cycle, highlighting the temporal evolution of the field patterns. Accompanying this supplementary material are videos showing the temporal evolution of the \(E_z\) field component, clearly visualizing the revolution of the field around the singularities. Two additional videos demonstrate that both the skyrmion density and the spin textures remain stable and robust throughout the pulse duration, emphasizing the inherent stability of these geometry-driven spin merons.

In the subsequent sections, we proceed to analyze the robustness and propagation characteristics of these spin textures in greater detail.

\section*{S3. Long-range integrity of the skyrmion tubes}

\begin{figure*}[htbp]
    \centering
    \includegraphics[width=\linewidth]{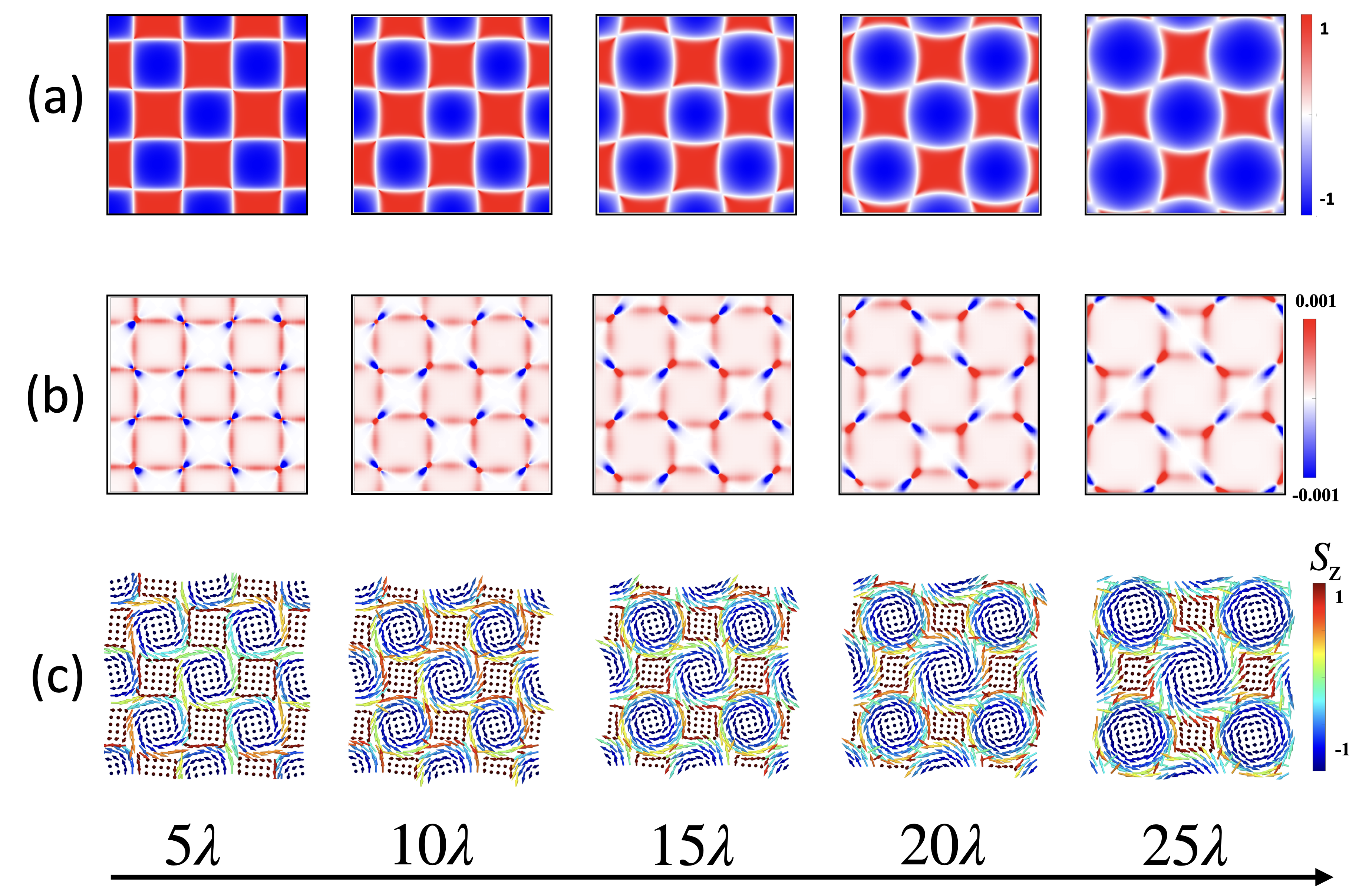}
    \caption{Propagation of the spin-meron lattice from \(z=5\lambda\) to
    \(25\lambda\).  
    Each column corresponds to the height indicated below.
    (a) Normalized longitudinal spin \(S_z/|\mathbf S|\); red denotes
    \(+1\), blue \(-1\).  
    (b) Skyrmion-density \(\rho_{\mathrm{sk}}\) (units \(10^{-3}\)).
    (c) 3-D spin texture: arrows give the in-plane spin direction,
    color encodes \(S_z\).  
    The axial (red) merons remain highly localized, while the central and
    diagonal (blue) merons expand and acquire additional azimuthal twist,
    yet the overall lattice and topological charges are conserved,
    confirming the existence of straight, robust skyrmion tubes at least
    up to \(25\lambda\).}
    \label{fig:S3}
\end{figure*}

To verify that the geometry-driven spin texture survives well beyond the
\(5\lambda\) range analyzed in the main text, we tracked the field out to a
distance of \(25\lambda\).  
Figure~\ref{fig:S3} collects the results at five monitor planes
(\(z=5\lambda,\,10\lambda,\,15\lambda,\,20\lambda,\,25\lambda\)). \textbf{Panel (a)} displays the normalized out-of-plane spin component \(S_z/|\mathbf S|\).  
The red (positive) axial merons on the cartesian axes remain sharply
defined at every height, whereas the blue central and diagonal merons
broaden gradually, evidencing a gentle diffraction-induced expansion
while retaining their sign and lattice registration.
\textbf{Panel (b)} shows the corresponding skyrmion-density maps
\(\rho_{\mathrm{sk}}\) (color bar \(10^{-3}\) normalized units).  
Even at \(25\lambda\) the alternating pattern of half-integer winding
numbers is clearly recognizable, confirming that the skyrmion charge of
each lattice site is conserved during propagation.
\textbf{Panel (c)} renders the in-plane spin vectors
(\(\mathbf S_\parallel/|\mathbf S|\), arrows) colored by \(S_z\).  
The central and diagonal Bloch merons exhibit an increasing azimuthal
twist—visible as tighter swirl patterns—yet the global square symmetry
and the axial Néel-like merons (red squares) remain intact.  
The persistence of all three features demonstrates that the
edge-imposed \(\pi/2\) phase ladder locks the vortex lattice so strongly
that the full 3-D spin-meron tubes remain coherent for at least
\(25\lambda\).

The continued persistence of both \(S_z\) and \(\rho_{\mathrm{sk}}\)
supports the Fresnel-number estimate of the main manuscript, suggesting
that the meron tubes should persist for hundreds of wavelengths before
paraxial spreading degrades the structure.

\section*{S4. Realistic silicon block: partial absorption and modified meron tubes}

For experimental feasibility it is essential that the effect persists in
common high-index photonic materials.  
We therefore replaced the perfectly absorbing square by a crystalline Si
block of identical lateral size (\(64\lambda\times64\lambda\)) and
thickness \(2\lambda\).  
At the design wavelength we take the complex index of silicon to be
\(n_\mathrm{Si}=3.48+0.006i\), resulting in \(\sim\!30\%\) absorption and
strong multiple reflections inside the block.

\begin{figure*}[htbp]
    \centering
    \includegraphics[width=\linewidth]{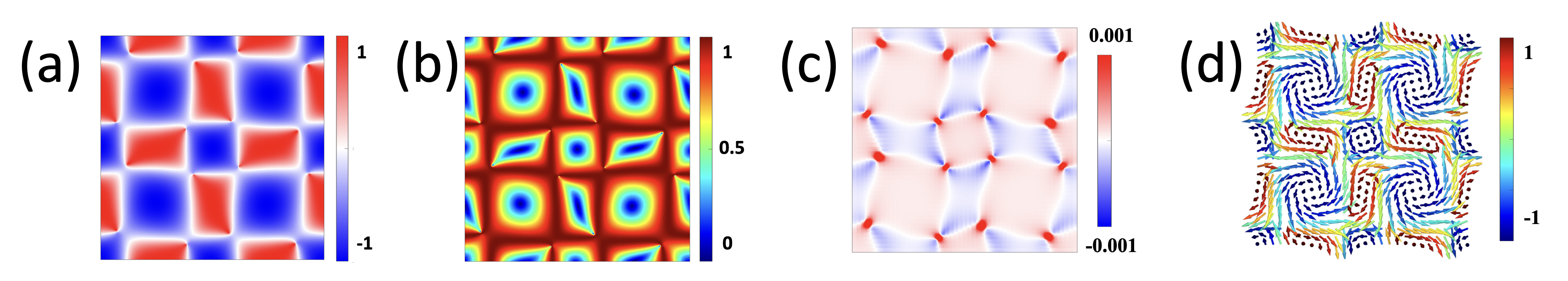}
    \caption{Spin-meron lattice generated by a realistic silicon block.
    (a) Normalized longitudinal spin component \(S_z/|\mathbf S|\).  
    (b) Normalized in-plane spin magnitude
    \(|\mathbf S_\parallel|/|\mathbf S|\).  
    (c) Skyrmion-density \(\rho_{\mathrm{sk}}\).  
    (d) 3-D spin texture: arrows denote the in-plane spin direction and
    colors represent \(S_z\).  
    Partial transmission and internal reflections in the high-index
    block distort the individual merons but leave their topological
    charges and overall lattice intact.}
    \label{fig:S4}
\end{figure*}

The partial transmission produces a richer interference pattern in the
diffracted field: waves leaked through the block overlap with those
emitted from the edges, slightly displacing the vortex cores and
distorting the local spin texture.  
Figure~\ref{fig:S4} summarizes the results at \(z=2\lambda\).

\begin{itemize}[leftmargin=*,nosep]
\item[(a)] Normalized longitudinal spin \(S_z/|\mathbf S|\).  
      The axial (red) and diagonal (blue) merons survive but their
      outlines are rounded into “pillows’’ and the central merons assume
      a lozenge shape.
\item[(b)] Normalized in-plane spin magnitude
      \(|\mathbf S_\parallel|/|\mathbf S|\).  
      The maxima now form skewed diamonds whose axes follow the local
      power-flow lines emerging from the partially transmitting faces.
\item[(c)] Skyrmion-density \(\rho_{\mathrm{sk}}\); despite the geometric
      distortions the alternating half-integer charge pattern is still
      clearly resolved.
\item[(d)] 3-D spin field: arrows give \(\mathbf S_\parallel/|\mathbf S|\)
      and color encodes \(S_z\).  
      The Bloch-type merons twist more strongly than in the ideal
      absorber, yet the global \(\mathcal C_4\) symmetry and the
      topological charges remain intact.
\end{itemize}

These observations confirm that the geometry-driven mechanism is
material-agnostic: even with significant internal reflections and only
partial absorption, a silicon block supports a stable lattice of
spin-meron tubes—albeit with deformed core shapes—thereby opening a
practical route to on-chip realization.

\bibliographystyle{apsrev4-1}

\bibliography{apssamp}